\documentclass[reprint,amsmath,amssymb,aps,prl,twocolumn,superscriptaddress,notitlepage,preprintnumbers]{revtex4-1}

\usepackage{graphicx}
\usepackage{dcolumn}
\usepackage{bm}
\usepackage{hyperref}

\usepackage{multirow}
\usepackage{enumitem}

\usepackage[english]{babel}
\usepackage{xcolor}
\hypersetup{
    colorlinks,
    linkcolor={red!80!black},
    citecolor={blue!50!black},
    urlcolor={blue!90!black}
}

\usepackage{color}
\usepackage{slashed,cancel}
\usepackage{xspace}
\newcommand{\ba} {\begin{equation}\begin{aligned}}
\newcommand{\ea} {\end{aligned}\end{equation}}
\newcommand{\be}{\begin{equation}}
\newcommand{\ee}{\end{equation}}

\newcommand{\cO}{\mathcal{O}}

\newcommand{\derp}{\partial}

\newcommand{\ov}[1]{\overline{#1}}

\newcommand{\GeV}{\ \text{GeV}}

\def\tH{\widetilde{H}}

\begin{document}

\preprint{IFT-UAM/CSIC-22-42}

\title{Right-handed neutrinos and the CDF II anomaly}

\newcommand{\IFT}
{Instituto de F\'isica Te\'orica UAM-CSIC, Calle Nicol\'as Cabrera 13-15, Cantoblanco E-28049 Madrid, Spain}

\newcommand{\UAM}
{Departamento de F\'isica Te\'orica, Universidad Aut\'onoma de Madrid, Cantoblanco, E-28049, Madrid, Spain}

\newcommand{\KTH}
{Department of Physics, School of Engineering Sciences, KTH Royal Institute of Technology, AlbaNova University Center, Roslagstullsbacken 21, SE-106 91 Stockholm, Sweden}

\author{Mattias Blennow}
\affiliation{\KTH}
\author{Pilar Coloma}
\affiliation{\IFT}
\author{Enrique Fern\'andez-Mart\'inez}
\affiliation{\IFT}
\affiliation{\UAM}
\author{Manuel Gonz\'alez-L\'opez}
\affiliation{\IFT}
\affiliation{\UAM}

\begin{abstract}
We point out that right-handed neutrinos can resolve the tension between the latest CDF-II measurement of the $W$-boson mass, $M_W$, and the standard model. Integrating out the new states yields a single $d=6$ operator, which induces a deviation from unitarity in the PMNS matrix. This alters the extraction of the Fermi constant from muon decay and increases the prediction for $M_W$, in line with the CDF~II result. Non-unitarity of the PMNS matrix would also affect beta, meson, and tau decays. We find that the CDF~II value for $M_W$ can be explained  without conflicting with lepton flavour universality constraints or the invisible decay width of the $Z$. However, the so-called Cabibbo angle anomaly is worsened if right-handed neutrinos are the origin of the $d=6$ operator. The situation improves if the operator coefficient is left unconstrained, implying additional sources of new physics, but a common explanation of both anomalies is in tension with universality bounds.
\end{abstract}

\maketitle

\textit{\textbf{Introduction.}}
The Standard Model of particle physics (SM) is an extremely successful theory, having made countless predictions that have been experimentally verified, the crowning achievement being the discovery of the Higgs boson a decade ago~\cite{ATLAS:2012yve,CMS:2012qbp}. With this in mind, any indications for physics beyond the SM (BSM) generally attract significant attention from the particle physics community. In particular, the so-called Cabibbo anomaly
~\cite{Seng:2018yzq,Czarnecki:2019mwq,Seng:2020wjq,Grossman:2019bzp} has attracted a lot of attention, and the recent release of the CDF II result for the mass of the W boson~\cite{CDFII}, including a 7 standard deviation discrepancy relative to the SM prediction, is certain to do the same. 

It would be particularly appealing to relate these budding new anomalies to longstanding open problems of the SM. The simplest extension of the SM able to accommodate the experimental evidence for neutrino masses and mixings, observed in neutrino oscillations, is arguably the addition of right-handed neutrinos to the SM particle content. If their Majorana masses (allowed by the SM gauge symmetry) are heavy, integrating them out yields the ($d=5$) Weinberg  operator~\cite{Weinberg:1979sa}, which generates the light neutrino masses after electroweak symmetry breaking. Following in the operator expansion, a single $d=6$ operator is generated at tree level~\cite{Broncano:2002rw}: 
\be
\cO_{d=6} = \ov{\ell_L}\tH\, c_{\cO_{d=6}}\,i\,\slashed{\derp}\left(\tH^\dag\ell_L\right)\,
\label{Genericd6Operator}
\ee
with
\be
c_{\cO_{d=6}}=Y_\nu\dfrac{1}{\Lambda^{2}}Y^\dag_\nu\,,
\label{coeff}
\ee
where $\Lambda$ is the Majorana mass of the right-handed neutrino and $Y_\nu$ is the matrix of Yukawa couplings  between the right-handed neutrinos, the left-handed lepton doublets $\ell_L$ and the Higgs $H$.

When the Higgs develops its vacuum expectation value, this $d=6$ operator induces a deviation from unitarity of the PMNS mixing matrix~\cite{Lee:1977tib,Shrock:1980vy,Schechter:1980gr,Shrock:1980ct,Shrock:1981wq,Langacker:1988ur,Bilenky:1992wv,Nardi:1994iv,Tommasini:1995ii,Bergmann:1998rg,Loinaz:2002ep,Loinaz:2003gc,Loinaz:2004qc,Antusch:2006vwa,Antusch:2008tz,Biggio:2008in,Alonso:2012ji,Abada:2012mc,Akhmedov:2013hec,Basso:2013jka,Abada:2013aba,Antusch:2014woa,Abada:2015oba,Abada:2015trh,Abada:2016awd}, which we will dub $N$~\cite{Fernandez-Martinez:2007iaa}:
\be
N_{\alpha i}=(\delta_{\alpha \beta}-\eta_{\alpha \beta})\,U_{\beta i}\,,
\ee
where $\eta = c_{\cO_{d=6}} v^2/4$ is a Hermitian \emph{positive semi-definite} matrix and $U$ is the unitary rotation that diagonalizes the Weinberg operator.

In the canonical type-I Seesaw mechanism~\cite{Minkowski:1977sc,Mohapatra:1979ia,Yanagida:1979as,Gell-Mann:1979vob}, where the lightness of neutrino masses derives from the hierarchy between the Dirac and Majorana masses, the $d=6$ operator will be highly suppressed, making its phenomenological impact negligible. However, naturally small neutrino masses may also arise from a symmetry argument. Indeed, the Weinberg operator is protected by the lepton number symmetry~\cite{Branco:1988ex,Kersten:2007vk,Abada:2007ux}, while the $d=6$ one is not. Thus, low-scale Seesaw variants with an approximately conserved lepton number symmetry are characterized by both naturally light neutrino masses and sizable Yukawa couplings, even for sterile neutrinos lying close to the electroweak scale. This is the case of the Inverse~\cite{Mohapatra:1986bd,Bernabeu:1987gr} or Linear~\cite{Malinsky:2005bi} Seesaw mechanisms.  In these scenarios the heavy neutrinos arrange in pseudo-Dirac pairs, with sizable mixing with the active neutrinos and significant unitarity deviations of the PMNS matrix~\cite{Langacker:1988ur,Antusch:2006vwa,Antusch:2014woa,Fernandez-Martinez:2016lgt}. All interactions involving neutrinos, both through charged and neutral currents, are consequently affected.

In this letter we discuss how these modifications affect the extraction of the Fermi constant in such a way that the prediction for the $W$-boson mass ($M_W$) shifts to larger values, reducing the tension with the new CDF II measurement. Conversely, the impact of the unitarity deviation in the extraction of $V_{ud}$ \emph{worsens} the so-called \emph{Cabibbo anomaly} as long as the coefficient of the $d=6$ operator in Eq.~(\ref{Genericd6Operator}) is positive semi-definite, as required by its generation from the inclusion of right-handed neutrinos. Nevertheless, if this assumption is relaxed and the coefficient of the operator is allowed to be \emph{indefinite}, it has been shown that a solution of the Cabibbo anomaly can also be found~\cite{Coutinho:2019aiy,Kirk:2020wdk}. This would however imply more elaborate additions to the SM particle content beyond only right-handed neutrinos (see Ref.~\cite{Coutinho:2019aiy} for a dedicated discussion). We will also relax this assumption when performing our global fit to the different relevant observables. In particular, we will show that bounds on lepton flavour universality are very important, and that the proposed $d=6$ operator can provide a very good fit to any two out of the three sets of observables: $M_W$ from CDF II, lepton flavour universality, and CKM unitarity. For the first two, a positive definite $\eta$ provides a good fit, and therefore only right-handed neutrinos are required for the ultraviolet (UV) completion. Interestingly, the common best fit is also in perfect agreement with the measurement of the invisible width of the $Z$. 

\textit{\textbf{Impact of non-unitarity on EW observables. }}
A non-unitary PMNS matrix would directly affect the dominant decay channel of the muon, from which the Fermi constant is extracted. This translates into the following relation between $G_F$ (the parameter which enters the Fermi Lagrangian) and $G_\mu$ (the value of the parameter extracted from the muon lifetime):
\be
G_F=G_\mu\left(1+\eta_{ee}+\eta_{\mu\mu}\right)\,.
\ee
As the prediction of the mass of the $W$ stems from the Fermi constant, the non-unitarity corrections in $G_F$ propagate to the prediction of $M_W$ as: 
\be
    M_W = M_Z \sqrt{\frac{1}{2}+\sqrt{\frac{1}{4}-\frac{\pi\alpha(1-\eta_{\mu\mu}-\eta_{ee})}{\sqrt{2}G_\mu M_Z^2(1-\Delta r)}}}\,,
\ee
where $\alpha$ is the fine-structure constant, $M_Z$ is the mass of the $Z$ gauge boson, and $\Delta r$ accounts for loop corrections.

Similarly, the invisible decay of the $Z$ is yet another observable, precisely measured, which is affected by non-unitarity. In particular, the number of active neutrinos extracted from this process is corrected to~\cite{Fernandez-Martinez:2016lgt}:
\be
N_\nu = 3-4\eta_{\tau\tau}-\eta_{ee}-\eta_{\mu\mu}\,.
\ee
Its value, measured at LEP~\cite{Janot:2019oyi}, constitutes a further source of information to constrain the entries of $\eta$. Notice that, disregarding the effect of $\eta_{\tau\tau}$, $N_\nu$ has the same functional dependence on the $\eta$ parameters as $M_W$. The resulting errors from the invisible width of the $Z$ are significantly larger and we will therefore not include this measurement in our fits. However, the current central value for $N_\nu$ is very compatible with the CDF~II result at just above a 1$\sigma$ difference in the resulting value of $\eta_{ee}+\eta_{\mu\mu}$.

If the $\eta$ matrix is positive semi-definite, its diagonal entries must be positive. Non-zero values of $\eta_{ee}$ and/or $\eta_{\mu \mu}$ would therefore increase the prediction for $M_W$. This would improve the agreement with the latest CDF~II measurement, pointing to a non-unitary PMNS matrix as a possible explanation to the strong tension with the SM. 

Nevertheless, these unitarity deviations are also strongly constrained by other observables. In particular, some of the strongest bounds arise from beta and kaon decays, from which the CKM matrix elements $V_{ud}$ and $V_{us}$ are extracted. The determination of $V_{ud}$ from superallowed beta decays would receive a non-unitarity correction with $\eta_{ee}$ entering the lepton vertex. This contribution is cancelled upon inclusion of $G_\mu$, since the same vertex is present in its determination, so that the final correction is:
\begin{equation}
\left|V_{ud}^{\beta}\right|=
\left(1+\eta_{\mu\mu}\right)\left|V_{ud}\right|,
\label{eq:betadec}
\end{equation}
where $V_{ud}^{\beta}$ is the experimentally measured value and $V_{ud}$ the actual entry of the CKM matrix.

The value of $V_{us}$ is determined through semi-leptonic kaon decays. The non-unitarity correction will in this case be different depending on the flavour of the final state lepton:
\begin{eqnarray}
\left|V_{us}^{K\rightarrow \pi e\overline{\nu}_{e}}\right|&=&
\left(1+\eta_{\mu\mu}\right)\left|V_{us}\right|,\\
\left|V_{us}^{K\rightarrow \pi \mu\overline{\nu}_{\mu}}\right|&=&
\left(1+\eta_{ee}\right)\left|V_{us}\right|.
\end{eqnarray}
These kind of semi-leptonic decays are controlled by a form factor, $f_+(q^2)$, which depends on the momentum transfer between the mesons. Experimental measurements are not able to disentangle $\vert V_{us}\vert$ from the form factor evaluated at zero momentum transfer, $f_+(0)$. An independent determination of the latter, arising from lattice QCD, is therefore needed.

While the new physics effects do affect the measurements of $V_{ud}$ and $V_{us}$, to alleviate the $\sim 3 \sigma$ tension of the Cabibbo anomaly, negative values of $\eta_{\mu \mu}$ would be required. As discussed in the introduction, this would imply a more complex extension of the SM than simply introducing right-handed neutrinos, but it is indeed a viable possibility~\cite{Coutinho:2019aiy} that will also be considered in our fit.

Finally, lepton flavour universality bounds also provide competitive limits on non-unitarity. These are derived from the relative branching ratios to different flavours, 
\be
R^P_{\alpha/\beta}=\Gamma(P\to\ell_\alpha \ov{\nu}_\alpha)/\Gamma(P\to\ell_\beta \ov{\nu}_\beta)\,,
\ee
so as to cancel uncertainties. Data coming from pion, kaon, and tau lepton decays constrain ratios of $\eta$ elements~\cite{Bryman:2021teu}:
\ba
\left(\frac{1-\eta_{\mu\mu}}{1-\eta_{ee}}\right)_\pi=&\,1.0010(9),\\
\left(\frac{1-\eta_{\mu\mu}}{1-\eta_{ee}}\right)_K=&\,0.9978(18),\\
\left(\frac{1-\eta_{\mu \mu}}{1-\eta_{e e}}\right)_\tau=&\,1.0018(14).\\
\ea

Notice that the CDF II measurement has a $\sim 3.6 \sigma$ tension with previous determinations of $M_W$. There are also several other measurements of $\sin \theta_W$ that constrain the same combination of elements of $\eta$ that appears in $M_W$ (see Refs.~\cite{Antusch:2014woa, Fernandez-Martinez:2016lgt}). These are all in agreement with the SM expectation, and hence in tension with the CDF~II determination of $M_W$. We will therefore not include them in our fit since the proposed scenario will not be able to modify that tension. Instead, we quantify the level of compatibility of observables sensitive to different combinations of elements of $\eta$. Furthermore, none of these other measurements have a sensitivity as good as that reported by the CDF
~II collaboration. Nevertheless, in case the new result does not pan out, we will also consider the current global determination~\cite{ParticleDataGroup:2020ssz} of $M_W = 80.379(12)$~GeV by other experiments, since it constrains the particular combination of elements of $\eta$ with higher accuracy than the alternative measurements of $\sin \theta_W$.  

Similarly, lepton flavour violating decays have also been shown to be excellent probes of the lepton mixing matrix unitarity deviations. However, these only constrain the \emph{off-diagonal} elements of $\eta$, which are not relevant to the present discussion. For example, very strong constraints on $\eta_{e \mu}$ exist from the non-observation of the lepton flavour violating $\mu \to e \gamma$ decay or $\mu \to e$ conversion in nuclei~\cite{Alonso:2012ji}. For a limited number of right-handed neutrinos, the structure of the $d=6$ operator can be constrained, or even derived, from that of the observed neutrino masses and mixings~\cite{Gavela:2009cd,Fernandez-Martinez:2015hxa}. Conversely, when three or more pseudo-Dirac pairs are considered, there are no constraints on the structure of $\eta$; in particular, $\eta_{e\mu}$ should only satisfy the Cauchy-Schwarz inequality $\eta_{e\mu}  \leq \sqrt{\eta_{ee} \eta_{\mu\mu} }$. Therefore, we will consider that $\eta_{ee}$ and $\eta_{\mu \mu}$ are not bounded by lepton flavour violating processes.
\begin{table}
\begin{center}
\renewcommand{\arraystretch}{1.8}
\begin{tabular}{|c|c|}
\hline
\textbf{Observable}&\textbf{Experimental measurement}\\
\hline
$\alpha$&$7.2973525693(11)\times10^{-3}$\\
\hline
$G_\mu$&$1.1663787(6)\times10^{-5}\GeV^{-2}$\\
\hline
$M_Z$&$91.1876(21)\GeV$\\
\hline
$M_W$ (PDG)&$80.379(12)\GeV$\\
\hline
$M_W$ (CDF II)~\cite{CDFII}&$80.4335(94)\GeV$\\
\hline
$\Delta r$&0.03652(22)\\
\hline
$N_\nu$~\cite{Janot:2019oyi}&2.9963(74)\\
\hline
$\vert V_{ud}\vert$&$0.97370(14)$\\
\hline
$\vert V_{us}\vert f_+(0)\,(K^\pm e3)$&$0.2169(8)$\\
\hline
$\vert V_{us}\vert f_+(0)\,(K^\pm \mu3)$&$0.2167(11)$\\
\hline
$\vert V_{us}\vert f_+(0)\,(K^L e3)$&$0.2164(6)$\\
\hline
$\vert V_{us}\vert f_+(0)\,(K^L \mu 3)$&$0.2167(6)$\\
\hline
$\vert V_{us}\vert f_+(0)\,(K^S e3)$&$0.2156(13)$\\
\hline
$f_+(0)$~\cite{Aoki:2021kgd}&$0.9698(17)$\\
\hline
\end{tabular}
\end{center}
\caption{List of relevant observables and their experimental determinations. All the values have been taken from Ref.~\cite{ParticleDataGroup:2020ssz} except explicitly stated otherwise. \label{tab:params}}
\end{table}

\textit{\textbf{Results. }}
In order to perform a fit to $M_W$ along with the results from beta and kaon decays, as well as with the universality constraints, we adopt a $\chi^2$ approach, where we add a Gaussian term for each of the observables listed in Tab.~\ref{tab:params}. Apart from the target parameters $\eta_{ee}$ and $\eta_{\mu\mu}$, the $Z$ boson mass $M_Z$, the value of $\Delta r$, the value of $f_+(0)$, and the true value of $V_{ud}$ were used as nuisance parameters, with the experimental bounds on the first three being introduced into the $\chi^2$ through pull terms. The true value of $V_{ud}$ was allowed to vary freely apart from the constraints introduced by the beta and kaon decays. The correlation matrix among the kaon decay observables from~\cite{Moulson:2017ive} has also been taken into account.

\begin{figure}
    \centering
    \includegraphics[trim=35 212 60 224,clip,width=0.45\textwidth]{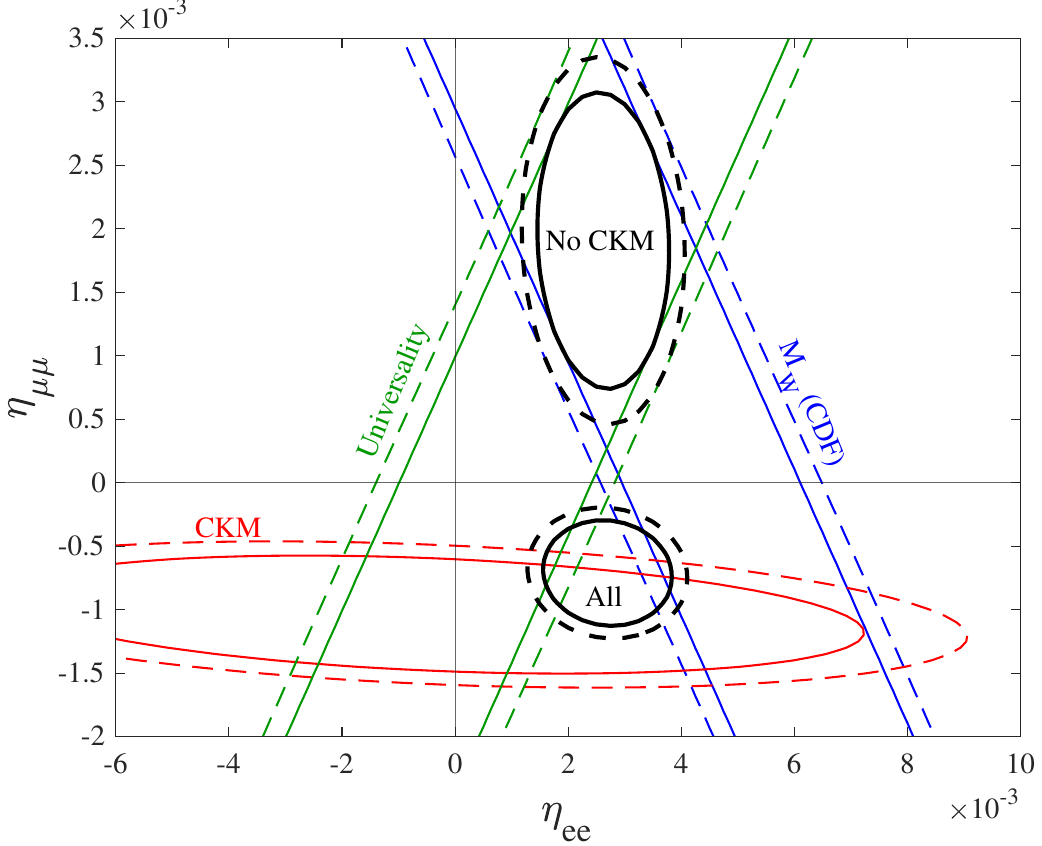}
    \includegraphics[trim=35 212 60 224,clip,width=0.45\textwidth]{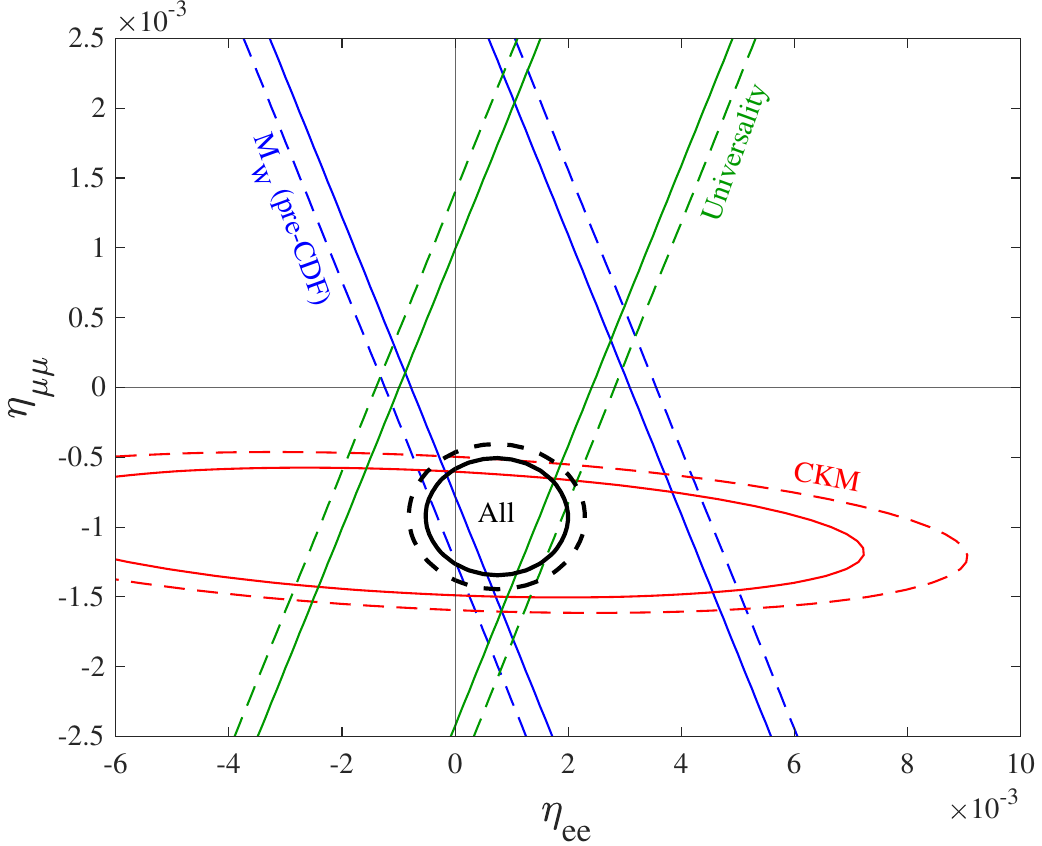}
    \caption{Results of our fit, projected onto the $\eta_{ee}$-$\eta_{\mu\mu}$-plane. The solid (dashed) curves correspond to the allowed regions at 95\% CL (99\% CL), for 2~d.o.f.. The constraints are labelled as `$M_W$' for the measurement of the $W$ mass, `Universality' for the lepton flavour universality bounds, and `CKM' for the bounds coming from beta and kaon decays. In the upper (lower) panel we take the constraints on $M_W$ from CDF~II~\cite{CDFII} (the global average before CDF~II~\cite{ParticleDataGroup:2020ssz}). The black contours correspond to combinations of different data sets, see text for details.}
    \label{fig:allowedanduniversality}
\end{figure}
In the upper panel of Fig.~\ref{fig:allowedanduniversality} we show the results for our fit to the $\eta$ coefficients. Our results show separately the preferred regions for the different sets of observables. As expected, the area in which the CDF-II measurement of $M_W$ would be reconciled with the prediction is very far from the SM, which corresponds to $\eta_{ee}=\eta_{\mu \mu}=0$. Similarly, the area in which the beta and kaon decay measurements would be reconciled with a unitary CKM matrix upon extraction of $V_{ud}$ and $V_{us}$ does not contain the SM expectation either. Finally, the bounds from the universality constraints prefer $\eta_{ee} \sim \eta_{\mu \mu}$. As can be seen, there is no common region where all three measurements overlap simultaneously. In fact, the global fit using all data sets is in some tension with all three sets and thus has its global minimum at a $\chi^2_{\rm min} / n_{\rm dof}= 40.1/7$, where $n_{\rm dof}$ is the number of degrees of freedom of the $\chi^2$. Nevertheless, the addition of new physics does provide a significant improvement with respect to the result obtained under the SM hypothesis: specifically, we find $\Delta \chi^2 =  \chi^2_{\rm SM} - \chi^2_{\rm min} = 48.1$ with $n_{\rm dof} = 2$.

It is therefore interesting to also consider scenarios in which one of the two present anomalies is not confirmed. In the lower panel of Fig.~\ref{fig:allowedanduniversality} we focus on the solution of the Cabibbo anomaly assuming that the $M_W$ measurement by CDF II will not be confirmed. The contours are therefore the same as for the upper panel except for the constraint from $M_W$, which now corresponds to the present global fit without CDF~II of $M_W = 80.379(12)$~GeV~\cite{ParticleDataGroup:2020ssz}. As can be seen, a satisfactory explanation of the Cabibbo anomaly can be found for positive values of $\eta_{ee}$ and negative values of $\eta_{\mu \mu}$, which is in agreement with previous results in Refs.~\cite{Coutinho:2019aiy,Kirk:2020wdk,Bryman:2021teu}. The best fit in this scenario has a $\chi^2/n_{\rm dof} = 9.7/7$. Notice that the preferred negative value for $\eta_{\mu \mu}$ foregoes the appealing simple explanation in terms of just right-handed neutrinos, which would require $\eta>0$, see Eq.~(\ref{coeff}).  

Conversely, in the upper panel of Fig.~\ref{fig:allowedanduniversality} we also show that the new $M_W$ anomaly from the CDF II measurement can be explained if the Cabibbo anomaly is not confirmed. This corresponds to the black contours labelled ``No CKM'', which provide a very good fit to both sets of observables and is also in excellent agreement with the invisible width of the $Z$. The best fit in this scenario has a $\chi^2/n_{\rm dof} = 3.3/2$.

\textit{\textbf{Summary and discussion. }}
In this letter we have explored the tantalizing possibility of a common explanation to the new $M_W$ measurement by CDF II and the Cabibbo anomalies, also linking them with the origin of neutrino masses and mixings by the inclusion of right-handed neutrinos in the particle spectrum. This induces a unique $d=6$ operator at tree level, whose coefficient $\eta$ may help to reconcile these measurements with predictions. We point out that, apart from these two sets of observables, the lepton universality constraints from the relative branching ratios of meson and tau lepton decays to different flavours are also very relevant to constrain the allowed parameter space. We find that the global fit including all constraints does offer a significant improvement over the SM only. Indeed, with the addition of only two parameters ($\eta_{ee}$ and $\eta_{\mu \mu}$), the global minimum of the $\chi^2$ is reduced by roughly 48 units when using all of the data sets. Nevertheless, significant tension between the three sets of observables remain, with a global minimum at $\chi^2/n_{\rm dof}=40.1/7$, corresponding to a $p$-value of only $1.3\cdot 10^{-6}$.

Thus, we also explored the possibility of explaining a single anomaly with the proposed scenario. When the CDF-II measurement of $M_W$ is not considered and replaced by the present determination of $M_W$ from other data, a very good fit to all observables is found for positive (negative) values of $\eta_{ee}$ ($\eta_{\mu \mu}$), confirming earlier analyses~\cite{Coutinho:2019aiy,Kirk:2020wdk,Bryman:2021teu}. Since the inclusion of right-handed neutrinos implies that $\eta$ should be positive semi-definite, this scenario would require a more complex UV completion~\cite{Coutinho:2019aiy}.

More interestingly, when assuming that a different explanation will be found for the Cabibbo anomaly and removing it from the fit, we obtained excellent fits to the $M_W$ measurement by CDF II in agreement with the lepton universality constraints as well as the invisible width of the $Z$. Furthermore, all diagonal elements of $\eta$ are positive in this case, in agreement with the type-I Seesaw expectation. They are also significantly different from 0, with the SM hypothesis disfavoured at $6.8 \sigma$, possibly linking the new determination of $M_W$ by CDF~II to the origin of neutrino masses. This interpretation of the new CDF II results implies $\eta_{ee} \sim \eta_{\mu \mu} \sim 3 \cdot 10^{-3}$, or, equivalently, a mixing of the heavy neutrinos with $\nu_e$ and $\nu_\mu$ of order $0.07$. For Yukawa couplings $\cO(1)$, the heavy neutrino mass would be around the TeV scale, within reach of future collider searches~\cite{Abdullahi:2022jlv}.

\textit{\textbf{Acknowledgements.}}
P.C., E.F.M., and M.G.L.\ acknowledge partial financial support by the Spanish Research Agency (Agencia Estatal de Investigaci\'on) through the grant IFT Centro de Excelencia Severo Ochoa No CEX2020-001007-S and by the grant PID2019-108892RB-I00 funded by MCIN/AEI/ 10.13039/501100011033, by the European Union's Horizon 2020 research and innovation programme under the Marie Sk\l odowska-Curie grant agreement No 860881-HIDDeN. P.C.\ is also supported by Grant RYC2018-024240-I, funded by MCIN/AEI/ 10.13039/501100011033 and by “ESF Investing in your future”.

\bibliographystyle{apsrev4-1}
\bibliography{references}

\end{document}